\documentclass[reprint, amsmath, amssymb, aps, superscriptaddress, preprintnumbers]{revtex4-2}
\usepackage{graphicx} 
\usepackage{epstopdf}
\usepackage{dcolumn}
\usepackage{bm}
\usepackage[colorlinks=true,linktocpage=true,linkcolor=blue,citecolor=blue,urlcolor=blue]{hyperref}
\usepackage{textcomp}
\usepackage{color}
\usepackage{multirow}
\usepackage{tabularx}
\usepackage{tabu}
\usepackage{makecell}

\def\gev{\rm GeV}

\begin{document}

\preprint{IFT-UAM/CSIC-25-74}

\title{Probing Top Quark - Electron Interactions at Future Colliders}

\def\BNL{High Energy Theory Group, Physics Department, 
    Brookhaven National Laboratory, Upton, NY 11973, USA}

\def\IFT{Departamento de F\'{i}sica Te\'{o}rica and Instituto de F\'{i}sica Te\'{o}rica UAM/CSIC, Universidad Aut\'{o}noma de Madrid, Cantoblanco, 28049, Madrid, Spain}

\author{Luigi Bellafronte}
\email{lbellafronte@fsu.edu}
\affiliation{Physics Department, Florida State University, Tallahassee, FL 32306-4350, USA}

\author{Sally Dawson}
\email{dawson@bnl.gov}
\affiliation{\BNL}

\author{Pier Paolo Giardino}
\email{pier.giardino@uam.es}
\affiliation{\IFT}

\author{Hongkai Liu}
\email{hliu6@bnl.gov}
\affiliation{\BNL}

\begin{abstract}
\noindent
Top quark interactions offer a window into possible new high scale physics and many models of new physics predict that the top quark interactions will deviate significantly from those predicted by the Standard Model. We present an analysis of the experimental restrictions on anomalous 4-fermion $e^+e^- t {\overline{t}}$ operators that is accurate to next-to-leading order (NLO) in both the electroweak and QCD interactions within the Standard Model Effective Field Theory framework. At NLO, there is sensitivity to an extended set of anomalous interactions beyond those probed at leading order. A comparison of current limits from electroweak precision observables, along with expected future limits from Drell-Yan and $t{\overline{t}}e^+e^-$ production at the high luminosity LHC, from deep inelastic scattering at the EIC, and from projected sensitivities at the future FCC-ee and CEPC machines demonstrates that each of these programs extends the precision understanding of the interactions of top quarks.

\end{abstract}
\maketitle
\section{Introduction and Motivation}

The top quark plays a special role in the Standard Model (SM). Because of its large mass,  the couplings of the top quark to the Higgs boson and to longitudinal gauge bosons are much larger than those of the light quarks, suggesting that the top quark may be fundamentally different from the lighter fermions.  Precision probes of top quark properties can thus help to elucidate questions of flavor in the quark sector and to determine whether there is new high scale physics affecting top quark interactions.

Limits on heavy new physics are often expressed in the framework of effective field theories (EFTs).   If the unknown new physics is much heavier than the weak scale, EFTs offer a consistent way to compare the capabilities of different current and future colliders.
By combining measurements from many sources, more stringent bounds are often obtained than are possible from a single experiment.  In this letter, we consider restrictions on operators involving the top quark in association with electrons from existing electroweak precision observables (EWPOs), neutral Drell-Yan and $pp \to t{\overline{t}}l^+l^-$ production at the high luminosity LHC (HL-LHC), polarized deep inelastic scattering (DIS) at the future Electron-Ion Collider (EIC)\cite{AbdulKhalek:2021gbh}, and from future $Z$-pole measurements, Higgs-Z associated  production and top-pair production at the proposed FCC-ee and CEPC $e^+e^-$ colliders \cite{FCC:2025lpp,Antusch:2025lpm,Altmann:2025feg,Celada:2024mcf}.  For each of these experimental measurements, we present the limits on anomalous top quark-electron interactions and combine our results into a global fit\footnote{We do not discuss the proposed ILC \cite{LinearColliderVision:2025hlt}, as we have not optimized for polarization.}.  We assume that the only anomalous interactions are 4-fermion operators with top quarks and leptons, and set all other non-SM interactions to zero.

Our study computes observables in the dimension-6 Standard Model Effective Field Theory (SMEFT) and is accurate to next-to-leading order (NLO) in both the QCD and electroweak sectors.  At NLO, there is sensitivity to new interactions that are not probed at lowest order (LO). Interactions with 4 light fermions are strongly restricted by current low energy data~\cite{deBlas:2013qqa,Falkowski:2017pss}.  
The two-fermion top-quark operators related to the vertices $t\bar t V$, $tbW$, and $t\bar t H$, along with the four-fermion top-quark operators $q\bar qt\bar t$ are also stringently constrained~\cite{Durieux:2018tev,Durieux:2018ggn,Durieux:2019rbz,Liu:2022vgo,Durieux:2022cvf,Severi:2022qjy,Cornet-Gomez:2025jot}. In this letter, we focus on the two-lepton-two-top-quark operators, which are currently only weakly constrained by the current LHC $pp\to t\bar t \ell^+ \ell^-$ measurements~\cite{CMS:2023xyc,ATLAS:2025yww}. 

One of the novel features of our study is that all relevant NLO electroweak corrections are included in the computations of observables.  NLO QCD corrections can be automated \cite{Brivio:2020onw,Degrande:2020evl}, but electroweak corrections must be computed on a case by case basis in the SMEFT.
The NLO electroweak contributions to Drell-Yan and DIS processes in the dimension-6 SMEFT from  4-fermion  top-electron operators are presented here for the first time. 

\section{SMEFT NLO Calculations}
The SMEFT assumes an SU(3)$_C\times$SU(2)$_L\times$U(1)$_Y$ gauge symmetry and contains only the SM particles. The Lagrangian is expanded around the SM Lagrangian,
\begin{equation}
	\mathcal{L}=\mathcal{L}_{SM}+\Sigma_{d,i}\frac{1}{ \Lambda^{d-4}} C_i^d O_i^d\, ,
\end{equation}
where $\Lambda$ is the scale of new physics and $O_i^d$ are the gauge invariant operators of dimension $d$ constructed out of SM particles and we restrict ourselves to dimension-6 operators.  The coefficient functions $C_i^d$ contain all of the information about potential new physics.  

The calculation of physical observables is both an expansion in powers of $\frac{1}{\Lambda}$ and in loops,
$\sim \frac{1}{16\pi^2}$ \cite{Buchalla:2022vjp}.  We present ${\cal{O}}(\frac{1}{16\pi^2 \Lambda^2})$ NLO SMEFT results that are accurate to one loop in the QCD and electroweak interactions using the dimension-6 SMEFT. 
The NLO results include both virtual one-loop contributions and the real emission of photons and gluons and are performed with dimensional regularization in $D=4-2\epsilon$ dimensions.  As input parameters, we take  
$M_W=80.369\pm .013~\gev$, $M_Z=91.1876\pm .0021~\gev$ , $G_F=1.1663787(6)\times 10^{-5}~/\gev^2$, 
$M_h=125.10\pm 0.14 ~\gev$,  
$M_t=172.9\pm 0.5~\gev$, and $\alpha_s(M_Z)=0.1181\pm 0.0011$. The CKM matrix is set to be diagonal and all fermions other than the top are assumed to be massless. We use a hybrid renormalization scheme:  SM parameters are renormalized on-shell, while coefficient functions, $C_i^d$ are renormalized using ${\overline{\rm MS}}$~\cite{Alonso:2013hga}.

Some of the processes we consider have been previously computed at NLO in the SMEFT:  the NLO EWPOs \cite{Dawson:2019clf,Bellafronte:2023amz,Biekotter:2025nln} and $e^+e^-\rightarrow Zh$ cross section \cite{Asteriadis:2024xuk,Asteriadis:2024xts} are known analytically. The NLO results for all dimension-6 SMEFT contributions except   from the 4-fermion interactions for Drell-Yan are known \cite{Dawson:2021ofa,Dawson:2018dxp}, while  the  4-fermion results for Drell-Yan and DIS are new in this work.

We use FeynRules \cite{feynrul} routines to convert  SMEFT $R_\xi$ Feynman rules \cite{Dedes:2017zog,Dedes:2023zws}  into a FeynArts \cite{feynarts} model file, then using FeynCalc \cite{feync}, we compute helicity amplitudes and reduce $1$-loop integrals to Passarino-Veltman integrals which are evaluated using Package-X/FeynHelpers~\cite{Patel:2015tea,Patel:2016fam,Shtabovenko:2016whf} and Looptools~\cite{Hahn:1998yk}. We do not fix the gauge in the SMEFT Lagrangian and we verify the cancellation of $\xi$ terms, in order to have a further check on our computations. 
To be consistent in the SMEFT expansion, our approach allows us to consider Feynman diagrams with at most a single SMEFT operator insertion. This ensures renormalizability by avoiding divergences that cannot be removed without introducing higher-dimension operators at leading order.

\begin{table}[htbp]
	\begin{tabular}{|c|c|}
		\hline
		\multicolumn{2}{|c|}{$\mathcal{O}^{(3),1133}_{\ell q}=(\bar \ell_L \gamma_\mu\tau^I \ell_L)(\bar Q_L\gamma^\mu\tau^IQ_L)$}  \\
		\hline
		$\mathcal{O}^{(1),1133}_{\ell q}=(\bar \ell_L\gamma_\mu \ell_L)(\bar Q_L\gamma^\mu Q_L)$ & $\mathcal{O}^{1133}_{\ell u} = (\bar \ell_L\gamma_\mu \ell_L)(\bar t_R\gamma^\mu t_R)$\\
		\hline
		$\mathcal{O}^{3311}_{qe} = (\bar e_R\gamma_\mu e_R)(\bar Q_L\gamma^\mu Q_L)$ &  $\mathcal{O}^{1133}_{eu} = (\bar e_R\gamma_\mu e_R)(\bar t_R\gamma^\mu t_R)$\\
		\hline
	\end{tabular}
	\caption{Dimension-6 operators containing 2 top quarks and 2 electrons that are considered in this study. $\ell_L$ is the first generation left-handed lepton doublet, $\ell_L^T=(\nu_{eL}, e_L)$, and  $Q_L^T\equiv  (t_L, b_L)$ is the third-generation left-handed quark doublet. $e_R$ and $t_R$ are the right-handed electron and top-quark, respectively.}
	\label{tab:opdef}
\end{table}

\begin{figure}
	\centering
	\includegraphics[width=0.35\columnwidth]{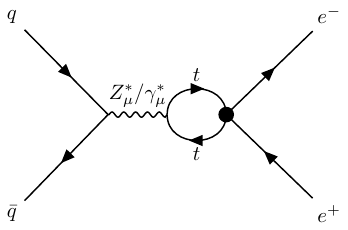}
	\includegraphics[width=0.35\columnwidth]{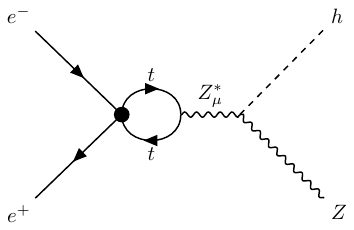}
	\includegraphics[width=0.23\columnwidth]{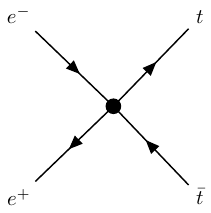}
	\caption{Representative Feynman diagrams for the Drell-Yan process $q\bar q\to e^-e^+$ (left), associated Z boson-Higgs production (middle), and top-pair production (right).}
	\label{fig:feyn}
\end{figure}

In this letter, we present complete NLO SMEFT results including the effects of all  two-electron-two-top  operators 
\begin{equation}
	\mathcal{O}^{(3),1133}_{\ell q}, \mathcal{O}^{(1),1133}_{\ell q}, \mathcal{O}^{3311}_{qe}, \mathcal{O}^{1133}_{\ell u}, \mathcal{O}^{1133}_{eu},
	\label{eq:4fops}
\end{equation}
where the four indices in the superscript denote the flavor of the four fermions \cite{Aguilar-Saavedra:2010uur}. Please note that we still refer to it as an NLO electroweak correction when a top quark appears in the loop, even though such contributions are leading order for the corresponding operators. The operators we consider are identical to those of the top-centric model of \cite{Bellafronte:2023amz}.
The operators in the Warsaw basis~\cite{Grzadkowski:2010es} are defined in Table~\ref{tab:opdef} and they contribute to the effective interactions $e^+e^-q\bar q$ (where $q$ is a light quark) and $e^+e^- Zh$ at NLO.  We do not consider the chirality flipped scalar and tensor operators as they do not interfere with the SM contribution in the limit of massless leptons. Nevertheless, those operators can be important in some precise measurements of the SM suppressed processes~\cite{Vasquez:2019muw}. Some sample Feynman diagrams are shown in Fig.~\ref{fig:feyn}. The DIS process is obtained by rotating the  Feynman diagram on the left-hand side by $90^o$.

\subsection{Drell-Yan}

\begin{figure}
	\centering
	\includegraphics[width=\columnwidth]{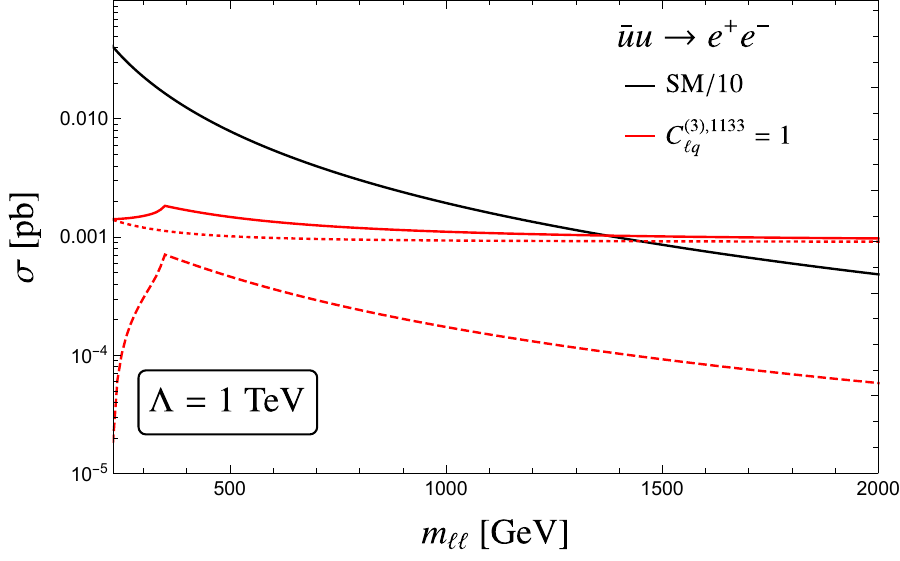}
	\caption{The partonic Drell-Yan cross section as a function of $\sqrt{{\hat s}} = m_{\ell\ell}$, comparing the SM prediction (black) with the contribution from the operator $\mathcal{O}^{(3),1133}_{\ell q}$ (solid red). The contribution from $\mathcal{O}^{(3),1133}_{\ell q}$ is separated into logarithmic (dashed red) and non-logarithmic (dotted red) terms.}
	\label{fig:DY}
\end{figure}

NLO QCD/electroweak results for neutral Drell-Yan production including all contributing operators except for four-fermion operators are given in \cite{Dawson:2021ofa}.  The four-fermion operators involving light quarks contribute at tree level \cite{deBlas:2013qqa} and are stringently limited from existing Drell-Yan measurements \cite{Farina:2016rws,Panico:2021vav,Torre:2020aiz}.  
The four-fermion operators involving electrons and top quarks defined in eq.~(\ref{eq:4fops})  contribute at NLO and analytic results for the helicity amplitudes are presented here for the first time and documented in the Supplemental Material~\cite{appendix}. 
In the limit where the partonic center of mass energy ${\hat{s}}\gg m_t^2$, the NLO SMEFT contributions to the  spin- and color-averaged partonic cross sections, ${\hat{\sigma}}^{\rm NLO}_{\rm DY} $, can be simplified, 
\begin{align}
	{\hat{\sigma}}^{\rm NLO}_{\rm DY} &= \frac{5 G_F^2}{324\Lambda^2\pi^3}\sum_{i}X_{0,i} C_i \nonumber\\
	&+ \frac{G_F^2}{162\Lambda^2\pi^3}\text{log}({\hat{s}}/m_t^2)\sum_i \left(\frac{m_t^2}{\hat{s}}\right)^{n_i} X_{1,i} C_i,
\end{align}
where $i$ runs over the operators of eq.~(\ref{eq:4fops}) and $C_i$ are the corresponding Wilson coefficients. $X_{0,i}$ and $X_{1,i}$ are functions of the weak boson masses and  $n_{i} =1$ for $\mathcal{O}^{(3)}_{\ell q}$ and $n_{i} =2$ for the remaining operators. The results are given in Table~\ref{tab:expr}. 
\begin{table*}[htbp]
	\begin{tabular}{|c|c|c|c|c|c|c|}
		\hline
		\multicolumn{2}{|c|}{Operators}  &  $\mathcal{O}^{(3),1133}_{\ell q}$ & $\mathcal{O}^{(1),1133}_{\ell q}$ &$\mathcal{O}^{1133}_{\ell u}$&$\mathcal{O}^{3311}_{qe}$& $\mathcal{O}^{1133}_{eu}$\\
		\hline\hline
		\multirow{2}{*}{$\bar u u\to e^+e^- $} & $(X_{0,i}) $& $\frac{w(2w +  z)}{4}$ & $-\frac{14 w^2 -31 wz +  17 z^2}{36}$ &$-\frac{14 w^2 -31 wz +  17 z^2}{18}$&$-\frac{17(w-z)^2}{18}$&$-\frac{17(w-z)^2}{9}$ \\\cline{2-7}
		&$ (X_{1,i})$ & $\frac{3(88 w^2-164 wz+85 z^2)}{16}$  & $-\frac{14 w^2-31 wz+17 z^2}{2}$  & $-\frac{32 w^2-22 wz+17 z^2}{8}$ & $-17(w-z)^2$ & $-\frac{8 w^2-25 wz+17 z^2}{4}$ \\\hline
		\multirow{2}{*}{$\bar d d\to e^+e^- $} & $(X_{0,i})$ & $\frac{w(4w-z)}{4}$  & $\frac{13 wz -8 w^2 -5 z^2}{36}$ & $\frac{13 wz-8 w^2- 5 z^2}{18}$ & $-\frac{5(w-z)^2}{18}$ & $-\frac{5(w-z)^2}{9}$\\\cline{2-7}
		&$ (X_{1,i})$  & $\frac{3(40 w^2-56 wz+25 z^2)}{16}$  & $-\frac{8 w^2-13 wz + 5 z^2}{2}$ & $-\frac{44 w^2-22 wz + 5 z^2}{8}$ & $-5(w-z)^2$ & $-\frac{14w^2-19 wz+5z^2}{4}$\\\hline
	\end{tabular}
	\caption{The analytic expressions for the NLO SMEFT functions $X_0$ and $X_1$ in the high-energy limit of $q {\overline{q}}\rightarrow e^+e^-$. Here, we define $w\equiv M_W^2$, $z\equiv M_Z^2$.}
	\label{tab:expr}
\end{table*}
The logarithmic terms are sub-dominant at the LHC due to the suppression of $(m_t^2/{\hat s})$ or $(m_t^2/{\hat s})^2$, and this is clear  in Fig.~\ref{fig:DY}, where we show the partonic SM cross sections (black curve) and the contributions from a representative SMEFT operator 
(solid red curve) as a function
of $\sqrt{\hat s}=m_{\ell\ell}$. The contributions from the logarithmic and non-logarithmic terms are shown by the dashed and dotted red curves, respectively. For illustration, we set $C^{(3),1133}_{\ell q}=1$, and the remaining coefficients to zero. Similar features hold for the other operators. The overall positive (negative) sign in $X_{0,i}$ indicates a constructive (destructive) interference with the SM. 

\subsection{Electron Ion Collider (EIC)}

By interchanging the Mandelstam variables, ${\hat s}\leftrightarrow {\hat t}$, the NLO results for polarized electron-quark scattering at the future EIC are obtained from the Drell-Yan results. The EIC is able to produce an electron beam with $P_e = 70\%$ polarization and a systematic uncertainty $\lesssim$1\%~\cite{AbdulKhalek:2021gbh} at $\sqrt{s} = 140$~GeV. To cancel the large SM photon contributions, we compute the left-right asymmetry
\begin{equation}
	A_{LR}\equiv (\tilde \sigma_L - \tilde\sigma_R)/(\tilde\sigma_L+\tilde\sigma_R)
\end{equation} 
as our observable at the EIC, where $\tilde\sigma_{L(R)} \equiv \sigma_{L(R)}(1+P_e)/2 + \sigma_{R(L)}(1-P_e)/2$, and $\sigma_{L(R)}$ is the hadronic cross section for the deep-inelastic scattering of a left (right)-handed electron.

\subsection{Electroweak Precision Observables}

The NLO SMEFT results for $Z$-pole observables were first computed in \cite{Dawson:2019clf}, and the results with an arbitrary flavor structure presented in \cite{Bellafronte:2023amz,Biekotter:2025nln}.  We include the following $Z$-pole observables,
\begin{equation}
	A_{\ell/b/s/c}, R_{\ell/b/c}, A_{\ell/b/s/c,\text{FB}},\sigma_h ,\Gamma_{Z}.
\end{equation}
The running of the electromagnetic coupling from $q^2=0$ to $q^2=M_Z^2$ and the W boson width, $\Gamma_{W}$, are additional precision observables that we include at NLO.

\subsection{FCC-ee}
The next run above the $Z$-pole at the FCC-ee is at the $WW$ threshold ($\sqrt{s}= 162$~GeV). The first three operators listed in eq.~(\ref{eq:4fops}) involving the left-handed quark doublet can modify the cross section of $b {\overline{b}}$ pair production. At LO at $\sqrt{s}= 162$~GeV,
\begin{align}
	&\sigma_{e^-e^+\to b \bar b}~(\rm pb)= 5.97-\biggl\{0.69 C^{3311}_{qe}\nonumber\\
	&-1.96 (C^{(3),1133 }_{\ell q}+C^{(1),1133}_{\ell q})\biggr\}\biggl(\frac{1~\textrm{TeV}}{ \Lambda}\biggr)^2.
\end{align}
The FCC-ee at $\sqrt{s}= 162$~GeV is expected to reach a precision of $2\times 10^{-4}$ on $\Delta R_b/R_b$~\cite{Greljo:2024ytg}.

The four-fermion electron-top operators contribute to Higgstrahlung at NLO as  depicted by the middle panel in Fig.~\ref{fig:feyn}, and the complete NLO SMEFT result for $\sigma(e^+e^-\rightarrow Zh)$ can be found in  ~\cite{Asteriadis:2024xuk,Asteriadis:2024xts}.

At $\sqrt{s}= 365$~GeV, the process $e^+e^-\rightarrow t {\overline{t}}$ becomes accessible and the electron-top four-fermion operators can be probed directly at LO as shown in the right panel of Fig.~\ref{fig:feyn}. 
The tree-level result at $\sqrt{s}= 365$~GeV is
\begin{align}
	&\sigma_{e^-e^+\to t \bar t}~(\rm pb)= 0.49+\biggl\{0.60 (C^{(3),1133 }_{\ell q}-C^{(1),1133}_{\ell q})\nonumber\\
	&-0.56 C^{1133}_{\ell u}-0.35 C^{3311}_{qe}-0.38 C^{1133}_{eu}\biggr\}\biggl(\frac{1~\textrm{TeV}}{ \Lambda}\biggr)^2.
\end{align}
The experimental uncertainty on the top-pair production cross section measurement at the FCC-ee with $2.7~\textrm{ab}^{-1}$ is expected to be 0.12\%~\cite{Defranchis:2025auz}. The theoretical uncertainty is estimated as 1.2\% by varying the top quark mass within $\pm 0.25$ GeV. We also include the forward-backward asymmetry of top quarks $A^t_{\rm FB}$, which can be calculated as
\begin{align}
	&A^t_{\rm FB}/10^3 = -88.3 + \biggl\{145.3 (C^{(1),1133 }_{\ell q}-C^{(3),1133}_{\ell q})\nonumber\\
	&-47.4 C^{1133}_{\ell u}-21.2 C^{3311}_{qe}+99.9 C^{1133}_{eu}\biggr\}\biggl(\frac{1~\textrm{TeV}}{ \Lambda}\biggr)^2.
\end{align}
The uncertainty is set to be 0.0016~\cite{Lroehrig2025} including both systematic and statistical ones. 

\section{Numerical Fits}

We are now in a position to perform a fit in which all processes are computed to NLO QCD and NLO electroweak order.  We emphasize that current global fits, while accurate to NLO QCD, typically do not include NLO electroweak corrections consistently.
``Partial'' NLO electroweak corrections were included in the global fits of \cite{Bartocci:2023nvp,Maura:2025rcv}, but the current calculation is the first to include NLO electroweak predictions for all observables, {\it{albeit}} for a restricted set of operators.

\begin{figure}
	\centering
	\includegraphics[width=0.9\columnwidth]{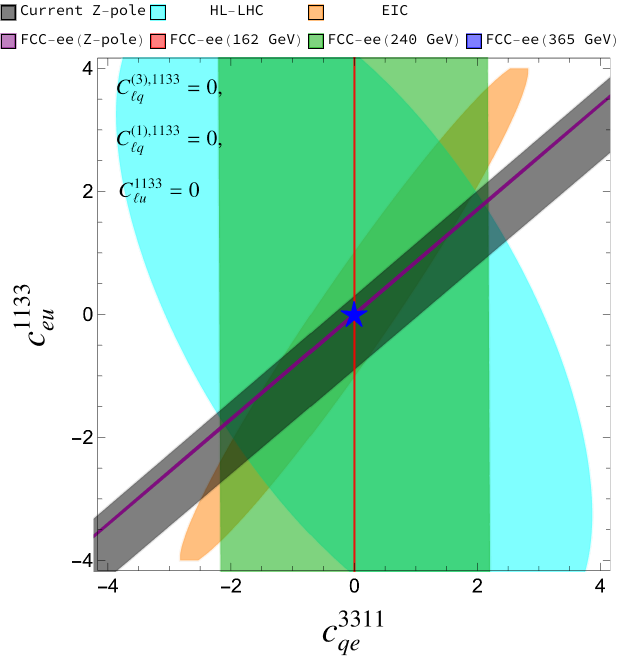}
	\includegraphics[width=0.88\columnwidth]{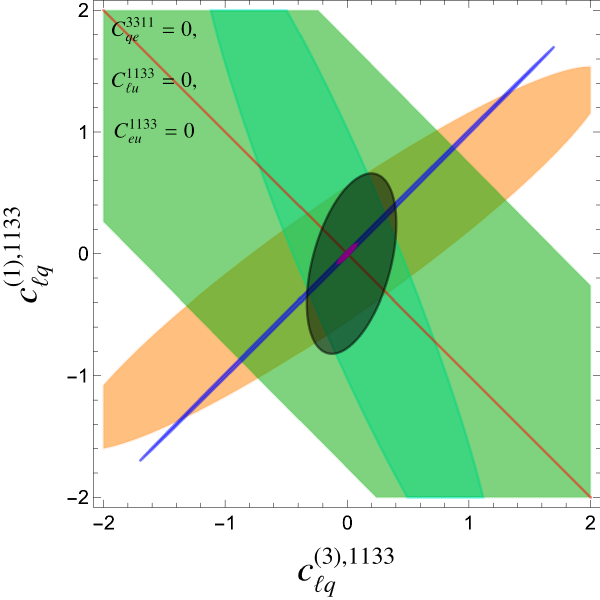}
	\caption{The $2\sigma$ limits on $C^{3311}_{qe}$-$C^{1133}_{eu}$ (upper panel) and $C^{(3),1133}_{\ell q}$-$C^{(1),1133}_{\ell q}$ (lower panel)  at current and future colliders. In the upper panel, the FCC-ee (365 GeV) projection appears as a dot at the origin (0,0). For clarity, we mark it with a blue star. All other operators are set to $0$ and $\Lambda=1$~TeV. The calculation does not include contributions of $\mathcal{O}(1/\Lambda^4)$.
	}
	\label{fig:cqe_ceu}
\end{figure}

We begin by illustrating the effects of different data sets to constrain the top quark-electron operators of eq.~(\ref{eq:4fops}) using a $\chi^2$ fit.  In Fig~\ref{fig:cqe_ceu}, we show the sensitivities of different future colliders to $C^{3311}_{qe}$-$C^{1133}_{eu}$ (upper panel) and $C^{(3),1133}_{\ell q}$-$C^{(1),1133}_{\ell q}$ (lower panel). All other coefficients are set to $0$.

Future $e^+e^-$ colliders
will operate at various center-of-mass energies, ranging from the $Z$-pole, the $WW$ threshold ($\sqrt{s}=$ 162 GeV) and the $Zh$ threshold ($\sqrt{s}=$ 240 GeV) to the top-pair threshold ($\sqrt{s}=$ 365 GeV). The future FCC-ee and CEPC are projected to produce trillions of $Z$ bosons at the $Z$-pole run, which will substantially improve the precision of $Z$-pole observables~\cite{Selvaggi:2025kmd,Maura:2024zxz}. We do a $\chi^2$ fit to EWPOs using the most precise theoretical results and the current experimental data, along with projections for the precision expected from FCC-ee $Z$-pole data.   
(The  data used in the fit are given in the Supplemental Material.)   The current limits from EWPOs are shown in Fig. \ref{fig:cqe_ceu} as the gray band, with the improvement from FCC-ee results in the purple band.

Higgstrahlung data have a different dependence on the top-quark electron operators from the EWPOs and the limit at $\sqrt{s}=240~\textrm{GeV}$ yields the green band\cite{Asteriadis:2024xts,Asteriadis:2024xuk,Maura:2025rcv}.  At $\sqrt{s}=365~\textrm{GeV}$, the Higgstrahlung results contribute to the fit, but the dominant contribution at this energy is from top quark pair production, which occurs at LO, shown as the dark blue band.
(The results for CEPC are similar to those of FCC-ee.  We do not include the ILC, as we have not optimized our study to include polarization effects relevant at the ILC.)

We adopt the HL-LHC Drell-Yan projections in~\cite{Greljo:2021kvv}. To ensure the validity of the EFT description, we restrict the Drell-Yan data to the 3 bins in the di-lepton invariant mass $m_{ee}$, which are [500,550], [550,650], and [650,800] GeV. The tree-level process $pp\to t\bar t e^+ e^-$ is also included in our analysis. We follow~\cite{tesina_abel} using the cross section measurements with four bins in $m_{ee}$, which are [100,120], [120,140], [140,180], and $>$180~GeV with 3 ab$^{-1}$ of data. Including both the NLO Drell-Yan and tree-level $pp\to t\bar te^-e^+$ contributions can break the degeneracies in the fit, as shown by the cyan line in Fig.~\ref{fig:cqe_ceu}. We have verified that the NLO contributions from the Drell-Yan process are numerically comparable with and complementary to the $pp\to t\bar te^-e^+$ contributions, even when considering only the low-mass $m_{\ell\ell}$ bins~\footnote{ See the supplementary material for Drell-Yan results including $m_{\ell\ell}$ bins up to 4 TeV.}.  In the Drell-Yan contribution, the tree-level process $b\bar b \to e^+ e^-$ is also included for the operators associated with the left-handed quark doublet. We find that the sensitivity to $C^{(3),1133}_{\ell q}/C^{(1),1133}_{\ell q}$ is dominated by the one-loop (tree-level) contribution. For $C^{(1),1133}_{\ell q}$ there is a cancellation between the up- and down-type quark  contributions at the NLO, as shown in Table~\ref{tab:expr}, while for $C^{3311}_{qe}$, the two contributions are comparable.
The first new data after the HL-LHC will come from deep inelastic scattering at the EIC.
We use 156 bins in the $x-Q^2$ plane with $Q^2 > 4~\text{GeV}^2$ and assume 500 fb$^{-1}$ integrated luminosity for each left-handed and right-handed electron beam mode leading to the sensitivity of the orange band. The renormalization/factorization scale is $\mu=\sqrt{Q^2}~(m_{\ell\ell})$ for the DIS (Drell-Yan) process.

\begin{figure}
	\centering
	\includegraphics[width=\columnwidth]{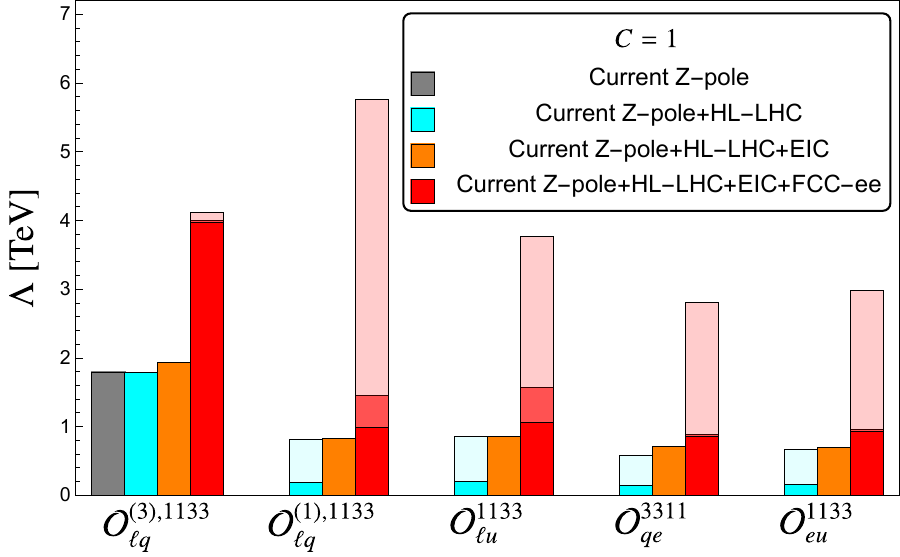}
	\caption{The 2$\sigma$ marginalized bounds on the scale $\Lambda$ from a global fit to the $4$-fermion operators considered in this work. The Wilson coefficients are set to 1. All contributions of $\mathcal{O}(1/\Lambda^4)$ are dropped. For the  HL-LHC, we consider only including the tree-level process $pp\rightarrow t \bar t e^- e^+$ (cyan) and including both the tree-level and NLO Drell-Yan processes (light cyan).
		The FCC-ee projections consist of three components, corresponding to FCC-ee $Z$-pole, FCC-ee at $WW$ threshold, and FCC-ee at $\sqrt{s}=365~\textrm{ GeV}$, shown in increasingly lighter shades of red. The contribution from FCC-ee at the $WW$ threshold is indistinguishable from the $Z$-pole contribution for operators $\mathcal{O}^{(3)}_{\ell q}$, $\mathcal{O}_{qe}$, and $\mathcal{O}_{eu}$. Higgstrahlung does not make an observable contribution to the FCC-ee bands.
	}
	\label{fig:bound}
\end{figure}

In Fig. \ref{fig:cqe_ceu}, we see that by combining different observables at a variety of energies in a global SMEFT analysis, the bounds can be significantly strengthened. However, the results from a fit to a single coefficient with the other operators marginalized over can give a quite different result from the single parameter fit, due to the presence of degeneracies.   In Fig. \ref{fig:bound}, we present a marginalized fit to the operators of eq.~(\ref{eq:4fops}) meant to demonstrate a possible sequence in time of bounds on the 
new physics scale $\Lambda$ (with the Wilson coefficients, $C_i$ set to 1).  With the current EWPO measurements, there is a limit 
of ${\cal {O}}(\sim 2~{\rm{TeV}})$ for $C^{(3)}_{\ell q}$, but there are degeneracies among the remaining contributions which do not allow for a meaningful bound. Including the current bounds from the measurement of $pp\to t\bar t e^+ e^-$ at the LHC~\cite{CMS:2023xyc,ATLAS:2025yww} can break the degeneracies, but does not yield meaningful results as the constraints on $C_i$ are beyond the perturbative limit at $\Lambda =1$~TeV.
Including the HL-LHC Drell–Yan and $pp \to t\bar te^+e^-$ projections effectively breaks the degeneracies in the remaining four-fermion operators, as illustrated by the cyan bands in Fig.~\ref{fig:bound}.
The orange bands in Fig.~\ref{fig:bound} show the bounds after adding the EIC data.  Finally, the results from the full FCC-ee program are shown as the red bands in Fig. \ref{fig:bound}.

Both the single-parameter and global-fit bounds on the Wilson coefficients (marginalized over the other 4 operators we consider)  with $\Lambda = 1$~TeV are shown in Table~\ref{tab:limit}.

\begin{table*}[htbp]
	\begin{tabular}{|c|c|c|c|c|c|c|}
		\hline
		\multicolumn{2}{|c|}{Observables} &  $C^{(3),1133}_{\ell q}$ & $C^{(1),1133}_{\ell q}$ &$C^{1133}_{\ell u}$&$C^{3311}_{qe}$& $C^{1133}_{eu}$\\
		\hline\hline
		\multirow{2}{*}{Current $Z$-pole} & single-parameter & [-0.21,0.32] &[-0.65,0.42] &[-0.36,0.55]&[-0.31,1.02]&[-0.87,0.26] \\ 
		\cline{2-7}
		& global-fit & [-0.29,0.31] & - & - & - & - \\ 
		\hline
		\multirow{2}{*}{Current $Z$-pole + HL-LHC ($t\bar te^-e^+$)} & single-parameter & [-0.21,0.32] &[-0.64,0.42] &[-0.36,0.55]&[-0.31,1.02]&[-0.85,0.26] \\ 
		\cline{2-7}
		& global-fit & [-0.29,0.31]  & [-30.7,24.6] & [-26.2,20.9] & [-40.2,51.4] & [-34.6,43.4] \\ 
		\hline
		\multirow{2}{*}{Current $Z$-pole + HL-LHC} & single-parameter & [-0.18,0.25] &[-0.53,0.37] &[-0.36,0.54]&[-0.31,0.97]&[-0.85,0.26] \\ 
		\cline{2-7}
		& global-fit & [-0.28,0.31]  & [-1.37,1.53] & [-1.37,1.39] & [-2.29,3.00] & [-2.27,2.21] \\ 
		\hline
		\multirow{2}{*}{Current $Z$-pole + HL-LHC+EIC} & single-parameter & [-0.17,0.23] &[-0.36,0.28] &[-0.30,0.43]&[-0.28,0.57]&[-0.66,0.26] \\ 
		\cline{2-7}
		& global-fit & [-0.27,0.24] &[-1.34,1.46] &[-1.33,1.38]&[-2.02,1.88]&[-2.10,1.46]  \\ 
		\hline
		\multirow{2}{*}{\makecell{Current $Z$-pole + HL-LHC+EIC\\+FCC-ee}} & single-parameter & [-0.0012,0.0012] &[-0.0012,0.0012] &[-0.0078,0.0078]&[-0.0033,0.0034]&[-0.016,0.015] \\ 
		\cline{2-7}
		& global-fit & [-0.058,0.059]  & [-0.029,0.030] & [-0.070,0.065] & [-0.12,0.13] & [-0.11,0.11] \\ 
		\hline
	\end{tabular}
	\caption{The 2$\sigma$ allowed regions on the five four-fermion electron-top operators with $\Lambda=1~{\textrm{TeV}}$. The global fit results are marginalized over the other two-electron-two-top-quark operators considered in this fit.}
	\label{tab:limit}
\end{table*}

\section{Conclusions}
In this letter, we studied the current and future constraints on  dimension-6 $4$-fermion operators involving both electrons and top quarks in the SMEFT framework.  The new feature of our study is that  QCD and electroweak NLO contributions are included consistently.

In a global fit to the top-electron 4-fermion operators, current $Z$-pole observables impose a stringent constraint on the SU(2)$_{L}$ triplet $4$-fermion operator $\mathcal{O}^{(3),1133}_{\ell q}$, with sensitivity reaching up to $\Lambda \sim 2$~TeV, while the other four top quark-electron operators are unconstrained due to degeneracies. Including the future HL-LHC tree-level $pp \to t\bar t e^+e^-$ projections breaks the degeneracies, leading to weak constraints. Adding the future NLO Drell-Yan measurements will significantly improve the limits on the Wilson coefficients of the four SU(2)$_{L}$ singlet $4$-fermion operators by more than an order of magnitude, as shown in Table~\ref{tab:limit}.  Finally, the FCC-ee $Z$-pole run is expected to reduce the experimental uncertainties on the $Z$-pole observables and has the potential to push the limit on $\Lambda$ for the operator $\mathcal{O}^{(3),1133}_{\ell q}$ to 4 TeV. Nonetheless, it has very limited effects on the other four SU(2)$_{L}$ singlet operators due to degeneracies. The full FCC-ee program is needed to probe the electron-top $4$-fermion operators at LO and to improve the limits on $\Lambda$ for the four SU(2)$_{L}$ singlet operators by a factor of $\sim 3 -6$.  We note that similar results can be obtained for CEPC.

\section*{Acknowledgements}

We thank Admir Greljo for pointing out the importance of the $R_b$ measurement above the $Z$-pole at the FCC-ee. P.P.G. is supported by the Ramón y Cajal grant~RYC2022-038517-I funded by MCIN/AEI/10.13039/501100011033 and by FSE+, and by the Spanish Research Agency (Agencia Estatal de Investigación) through the grant IFT Centro de Excelencia Severo Ochoa~No~CEX2020-001007-S. The work of LB is supported in part by the U.S.
Department of Energy under Grant No. DE-SC0010102 and by the College of Arts and Sciences of Florida State
University. S.D. and H.L.  are supported by the U.S. Department of Energy under Grant Contract~DE-SC0012704. Digital data is provided in the supplemental files.
\bibliographystyle{apsrev4-1}
\bibliography{eic.bib}
\clearpage
\onecolumngrid

\setcounter{equation}{0}
\setcounter{figure}{0}
\setcounter{table}{0}
\setcounter{section}{0}
\makeatletter

\begin{center}
 \large{\bf 
Probing Top Quark Interactions at Future Colliders
 }\\
 Supplemental Material
\end{center}

In this supplemental material, we present the numerical values used in the $Z$-pole observable analysis in Table~\ref{tab:ewpo}.
\begin{table}[h]
\begin{center}
\begin{tabular}{|l|c|c|c|}
\hline
Observables & Experimental values & current (future) theoretical values & FCC-ee projections
\\
\hline\hline
$1/\alpha_{\rm QED}$  & $128.952\pm 0.014$& $ 128.316\pm 0.009$ & $\pm 0.8\times 10^{-3}$\\\hline
$\Gamma_W$(GeV)  & $2.085\pm 0.042$& $ 2.0903\pm  0.0003$\cite{Cho:2011rk} & $\pm2.7\times 10^{-4} \pm 2.0\times 10^{-4}$\\\hline
$\Gamma_Z$(GeV) & $2.4955\pm 0.0023$ & $2.4943\pm \textcolor{black}{0.0004}~(8.0 \times 10^{-5})$\cite{Freitas:2014hra,Dubovyk:2018rlg,Dubovyk:2019szj}  &  $\pm4\times10^{-6}\pm 12\times10^{-6}$
\\\hline
$R_e$ & $20.804\pm 0.05$ &  $20.732\pm \textcolor{black}{0.006}~(1.2\times 10^{-3}) $\cite{Freitas:2014hra,Dubovyk:2018rlg,Dubovyk:2019szj} & $\pm3.4\times10^{-6}\pm 2.3\times10^{-6}$ \\
\hline
$R_\mu$ & $20.784\pm 0.034$ & $20.732\pm \textcolor{black}{0.006}~(1.2\times 10^{-3}) $\cite{Freitas:2014hra,Dubovyk:2018rlg,Dubovyk:2019szj} &  $\pm2.4\times10^{-6}\pm 2.3\times10^{-6}$ \\
\hline 
$R_\tau$ & $20.764\pm 0.045$ & $ 20.779\pm \textcolor{black}{0.006}~(1.2\times 10^{-3})$\cite{Freitas:2014hra,Dubovyk:2018rlg,Dubovyk:2019szj} &  $\pm2.7\times10^{-6}\pm 2.3\times10^{-6}$ \\
\hline 
$R_b$& $0.21629\pm 0.00066$ & $ 0.2159\pm 0.0001~(2\times 10^{-5})$\cite{Freitas:2014hra,Dubovyk:2018rlg,Dubovyk:2019szj}&  $\pm1.2\times10^{-6}\pm 1.6\times10^{-6}$ \\
\hline
$R_c$ & $0.1721 \pm 0.0030$ & $ 0.1722\pm 0.00005~(1\times 10^{-5})$\cite{Freitas:2014hra,Dubovyk:2018rlg,Dubovyk:2019szj} &  $\pm1.4\times10^{-6}\pm 2.2\times10^{-6}$ \\
\hline
$\sigma_h$(nb) & $41.481\pm 0.033 $ & $41.492\pm \textcolor{black}{0.006}~(5\times 10^{-4})$\cite{Freitas:2014hra,Dubovyk:2018rlg,Dubovyk:2019szj}& $\pm3\times10^{-5}\pm 8\times10^{-4}$ \\
\hline 
$A_e ({\text{from}}~A_{LR}~{\text{had}}) $ & $0.15138\pm 0.00216$ & $ 0.1469\pm 0.0004$\cite{Dubovyk:2019szj,Awramik:2006uz} & \multirow{3}{*}{$\pm1.4\times 10^{-5}$}\\
\cline{1-3}
$A_e ({\text{from}}~A_{LR}~{\text{lep}})$ & $0.1544\pm 0.0060$ & $ 0.1469\pm 0.0004$\cite{Dubovyk:2019szj,Awramik:2006uz} & \\
\cline{1-3}
$A_e ({\text{from~Bhabba~pol}})$ & $0.1498\pm 0.0049$ & $ 0.1469\pm 0.0004$\cite{Dubovyk:2019szj,Awramik:2006uz} & \\
\hline
$A_\mu$ & $0.142\pm 0.015$ & $ 0.1469\pm 0.0004$\cite{Dubovyk:2019szj,Awramik:2006uz} & $\pm3.2\times 10^{-5}$\\
\hline
$A_\tau ({\text{from~SLD}})$ & $0.136\pm 0.015$ & $ 0.1469\pm 0.0004$\cite{Dubovyk:2019szj,Awramik:2006uz} & \multirow{2}{*}{$\pm3.4\times 10^{-5}$}\\
\cline{1-3}
$A_\tau (\tau~ {\text{pol}})$ & $0.1439\pm 0.0043$ & $ 0.1469\pm 0.0004$\cite{Dubovyk:2019szj,Awramik:2006uz} &  \\
\hline
$A_c$ & $0.670\pm 0.027$ & $ 0.66773\pm 0.0002$\cite{Dubovyk:2019szj,Awramik:2006uz} & $\pm6.0\times 10^{-5}$\\
\hline 
$A_b$ & $0.923\pm 0.020$ & $0.92694\pm 0.00006$\cite{Freitas:2014hra,Dubovyk:2018rlg,Dubovyk:2019szj} & $\pm9.8\times 10^{-5}$
\\ \hline  
$A_s$ & $0.895\pm 0.091$ & $  0.93563\pm 0.00004$\cite{Dubovyk:2019szj,Awramik:2006uz} & $\pm1.2\times 10^{-4}$\\
 \hline  
$A_{e,FB}$ & $0.0145\pm 0.0025$ & $ 0.0162\pm 0.0001$\cite{Dubovyk:2019szj,Awramik:2006uz} & $\pm3.3\times 10^{-6} \pm 2.4\times 10^{-6}$ \\
 \hline
$A_{\mu,FB}$ & $0.0169\pm 0.0013$ & $ 0.0162\pm 0.0001$\cite{Dubovyk:2019szj,Awramik:2006uz} & $\pm2.3\times 10^{-6} \pm 2.4\times 10^{-6}$ \\
 \hline
 $A_{\tau,FB}$ & $0.0188\pm 0.0017$ & $ 0.0162\pm 0.0001$\cite{Dubovyk:2019szj,Awramik:2006uz} & $\pm2.8\times 10^{-6} \pm 2.4\times 10^{-6}$  \\
 \hline
$A_{b,FB}$ & $0.0996\pm 0.0016$ & $ 0.1021\pm 0.0003$\cite{Freitas:2014hra,Dubovyk:2018rlg,Dubovyk:2019szj} & $\pm4\times 10^{-6} \pm 4\times 10^{-6}$ \\
\hline
$A_{c,FB}$ & $0.0707\pm 0.0035$ & $ 0.0736\pm 0.0003$\cite{Dubovyk:2019szj,Awramik:2006uz} & $\pm5\times 10^{-6} \pm 5\times 10^{-6}$\\
\hline
$A_{s,FB}$ & $0.0976\pm 0.0114$ & $0.10308 \pm 0.0003$\cite{Dubovyk:2019szj,Awramik:2006uz} & $\pm7.4\times 10^{-6} \pm 7.4\times 10^{-6}$ \\
\hline
$M_W $(GeV)~{\text{PDG~World~Ave}} & $80.377\pm 0.012$ &$ 80.357\pm \textcolor{black}{0.004}~(0.001)$\cite{Awramik:2003rn,Erler:2019hds} & $\pm1.8\times 10^{-4} \pm 1.6\times 10^{-4}$\\ 
\hline
\end{tabular}
\caption{The numerical values used in the $Z$-pole observables analysis. The experimental results are taken from the Particle Data Group\cite{ParticleDataGroup:2018ovx}. The FCC-ee projections are taken from~\cite{Selvaggi:2025kmd}. Please note that the theoretical central value of $1/\alpha_{\rm QED}$ is obtained at one-loop level, while in our analysis, we set it equal to the experimental central value. }
\end{center}
\label{tab:ewpo}
\end{table}

\begin{figure}
	\centering
	\includegraphics[width=\columnwidth]{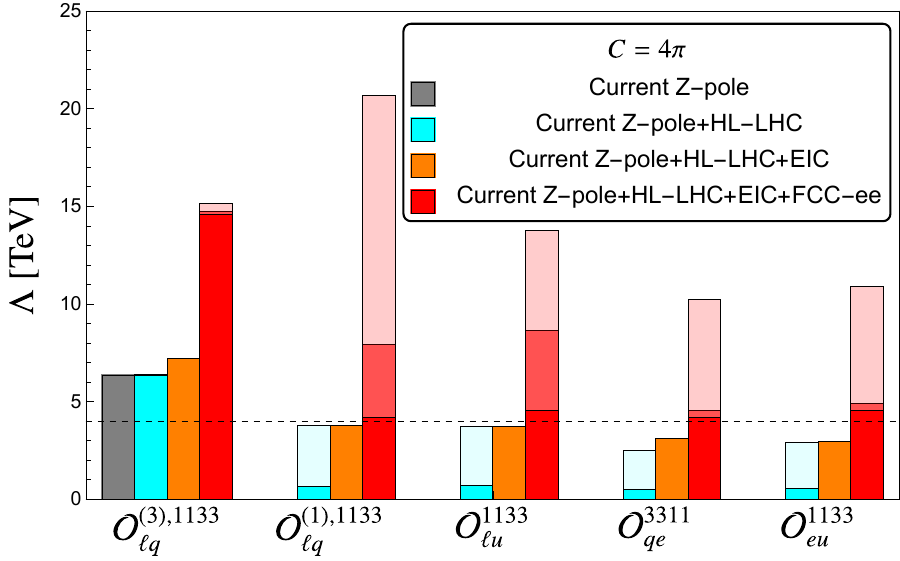}
	\caption{Same as Fig.~\ref{fig:bound} in the main text, except the Wilson coefficients are set to $4\pi$.}
	\label{fig:bound_high}
\end{figure}

In Fig.~\ref{fig:bound_high}, we show the bounds on $\Lambda$, assuming $C = 4 \pi$ and using the $m_{\ell\ell}$ bins from 500 GeV to 4 TeV. It shows that for the four SU(2)$_L$ singlet operators, the bounds on $\Lambda$ from the HL-LHC are less than 4 TeV, indicating the inconsistency of using $m_{\ell\ell}$ bins up to 4 TeV.

\section{readme}
{\bf{Amplitudes for Drell-Yan}}

This folder contains amplitudes for the ${\overline {u}} u \rightarrow  e^+ e^-$, ${\overline {d}} d \rightarrow e^+ e^-$, ${\overline {u}} u\rightarrow \mu^+\mu^-$ and ${\overline {d}} d \rightarrow \mu^+\mu^-$  processes  for the operators discussed in this paper.

We note that file names use "T" instead of "C" for the Wilson coefficients. All files are in txt format.

{\bf{Results}}

The results are  presented in the form, for example:
$$T_{qe}[3,3,1,1].{\textrm{txt}}\equiv A_{NLO, T_{qe}[3,3,1,1]} 
$$
where $A_{NLO, T_{qe[3,3,1,1]}} $ contains the chirality amplitudes proportional to $C_{qe}^{3311}$  and the flags $ll, rr, rl$ and $lr$= 1,0 separate the different chiralities.  (Note that the relevant coefficients are contained in the files.) Color delta for incoming quarks must be included for phenomenology.
The finite parts of the 1-loop amplitudes have been obtained by subtracting infrared poles (indicated as IR in the folder).

{\bf{Notation}}

\begin{itemize}
	
	\item
	$s_1$, $t_1$, $u_1$: Mandelstam variables
	
	\item
	$z$, $w$, $h$, $t$: masses ${\bf{squared}}$ of the Z, W, Higgs and Top respectively
	
	\item
	$v$: Higgs vacuum expectation value ${\bf{squared}}$.  To connect it to the  physical value one has to write 
	$$
	v \rightarrow \frac{1}{\sqrt{2} G_F} \left(1 + \frac{1}{2 G_F\Lambda^2} (\sqrt{2} (C_{fl}^{(3)}[1,1]+C_{fl}^{(3)}[2,2]) - \frac{1}{\sqrt{2}} (C_{ll}[2, 1, 1, 2]+C_{ll}[1, 2, 2, 1]))\right) \, 
	$$
	
	\item Lam : $\frac{1}{\Lambda^2}$ where Lambda is the scale of new physics 
	\item $\mu$ is renormalization scale ${\bf{squared}}$
	
	\item 
	a0f, b0f, c0f,  and d0f are the finite parts of the PV integrals $A_0[\dots]$, $B_0[\dots]$, $C_0[\dots]$ and $D_0[\dots]$
	
	\item
	Fermionic operators include the dependence on the flavor structure in the name. For example $[2,2]$ in $C_{fl}^{(3)}[2,2]$ represents two insertions of second-generation fermions.
\end{itemize}

Some operators have been redefined to eliminate square roots, the translation is as follows:
\begin{itemize}
	\item 
	$C_{Wr} \rightarrow \sqrt{\frac{w}{v}} \cdot C_W$
	\item
	$C_{\phi WBr} \rightarrow \sqrt{\frac{w}{z - w}} \cdot C_{\phi WB}$
	\item
	$C_{uBr}[3, 3] \rightarrow \sqrt{\frac{2 t}{z - w}} \cdot C_{uB}[3, 3]$
	\item $C_{uWr}[3, 3] \rightarrow \sqrt{\frac{2 t}{w}} \cdot C_{uW}[3, 3]$
\end{itemize}

{\bf{Flags}}

$ll$, $rr$, $rl$, $lr$: Chirality flags, for example rl refers to the  chirality *right* of the incoming particles and the  chirality *left* of the outgoing particles .

{\bf{Cross Section}}

To find the spin- and color-averaged cross section define for example,
$$A_{NLO,Tqe[3,3,1,1] }^{LL}=A_{NLO, Tqe[3,3,1,1]}\, , 
(ll=1,rr=0,rl=0,lr=0)\, .$$
The spin and color-averaged differential cross section of the hard scattering $\bar q(k_1) q(k_2) \to e^+(p_1) e^-(p_2)$ is
$$\frac{d\bar\sigma_{NLO}}{dt_1} = \frac{1}{16\pi s_1^2}\frac{1}{12}\sum_{X,Y} |\sum_{i}A_{NLO,i}^{XY}|^2|\hat M_{XY}|^2$$
where $s_1 = (k_1+k_2)^2$, $t_1 = (k_1-p_1)^2$, and $\hat M_{XY}$ are the stripped helicity amplitudes, with $X$ and $Y$ representing the helicity indices
$$|\hat M_{LL}|^2 = |\hat M_{RR}|^2 = 4 u_1^2,\quad|\hat M_{LR}|^2 = |\hat M_{RL}|^2 = 4 t_1^2.$$

\end{document}